\documentclass{jps-cp}
\usepackage{txfonts} 
\usepackage{wrapfig}
\usepackage{braket}
\title{Magnetic anisotropy of chiral magnet Yb(Ni$_{1-x}$Cu$_x$)$_{3}$Al$_{9}$ \\
at high magnetic fields}

\author{Daichi \textsc{Ito}$^{1}$ , Takeshi \textsc{Matsumura}$^{1}$ and Shigeo \textsc{Ohara}$^{2}$}

\inst{$^{1}$Department of Quantum Matter, AdSM, Hiroshima University, Higashi-hiroshima
739-8530, Japan \\
$^{2}$Department of Physical Science and Engineering, Graduate School of Engineering, Nagoya Institute of Technology, Nagoya 466-8555, Japan}

\email{m184185@hiroshima-u.ac.jp}

\recdate{September 14, 2019}

\abst{\quad We have measured the magnetization of rare-earth chiral magnets Yb(Ni$_{1-x}$Cu$_x$)$_{3}$Al$_{9}$ for $x$=0 and 0.06 up to 14.5\,T. Magnetization for the field direction $\sl{H}$\,$||$\,$\sl{c}$, which is the hard axis at low temperatures and low fields, overtakes the easy axis magnetization for $\sl{H}$\,$\perp$\,$\sl{c}$ above $\sim$\,4\,T.
We analyzed this overtaking of the magnetization curves by considering a crystalline electric field model. 
It is shown that the $\sl{B}_{66}$ term, which connects the $\ket{\pm5/2}$ and the $\ket{\mp7/2}$ states through the off-diagonal matrix element, is responsible for the characteristic magnetization process in this compound. We also introduce orbital dependent exchange interactions to explain the temperature dependence of magnetic susceptibility.}

\kword{YbNi$_3$Al$_9$, mgnetization, crystal electric fields, chiral soliton lattice, rare earth compounds, Dzyaloshinskii-Moriya(DM) interection}

\begin{document}
\maketitle

\section{Introduction}

YbNi$_3$Al$_9$ crystallizes in an ErNi$_3$Al$_9$-type structure with the space group $\sl{R}$32\,(No.155), where both inversion and mirror symmetries are broken\cite{1}.
The magnetic moments of the Yb$^{3+}$ ions order at $\sl{T}$$_{\rm{N}}$\,=\,3.5\,K, where the moments lie in the hexagonal $\sl{c}$-plane and propagate along the $\sl{c}$-axis with an incommensurate wave vector, indicating that the magnetic structure is helical\cite{2,3}. 
When a magnetic field is applied along the $\sl{c}$-plane, a transition to the ferromagnetic state occurs at $\sl{H}$$_{\rm{c}}$\,=\,1\,kOe. 
Interestingly, by substituting Cu for Ni, both $\sl{T}$$_{\rm{N}}$ and $\sl{H}$$_{\rm{c}}$ increases. 
At $x$\,=\,0.06, a characteristic $\sl{M}$($\sl{H}$) curve is observed, which is reminiscent of a chiral soliton lattice\,(CSL) state, a periodic array of incommensurate chiral spin twist formed in magnetic fields applied perpendicular to the helical axis in monoaxial chiral helimagnets\cite{4,5}.
CSL is a nonlinear order of topological spin structure originating from the crystal chirality through an antisymmetric interaction, the so-called Dzyaloshinskii--Moriya\,(DM) interaction.
By resonant x-ray diffraction, we have recently shown that the helical magnetic structure of Yb(Ni$_{1-x}$Cu$_x$)$_{3}$Al$_{9}$ has a fixed sense of spin rotation corresponding to the crystal chirality and have confirmed that the CSL state is formed in magnetic fields\cite{6}.
This series of $\sl{f}$-electron compounds with the ErNi$_3$Al$_9$-type structure have been attracting interest as chiral magnets where new type of ordered structure can be realized\cite{7,8,9,10,11}.\par
Although the hexagonal $\sl{c}$-plane is considered to be an easy plane of magnetization both from the helimagnetic structure and from the ferromagnetic transition above $\sl{H}$$_{\rm{c}}$ for $\sl{H}$\,$\perp$\,$\sl{c}$, the $\sl{M}$($\sl{H}$) curves show that the $\sl{c}$-axis becomes the easy axis of magnetization at high fields.
The $\sl{M}$($\sl{H}$) curve for $\sl{H}$\,$||$\,$\sl{c}$ overtakes the $\sl{M}$($\sl{H}$) curve for $\sl{H}$\,$\perp$\,$\sl{c}$ at around 4\,T\cite{2}.
Also in the temperature dependence of magnetic susceptibility, $\chi$$_{\sl{H}\,||\,\sl{a}}$ is larger than $\chi$$_{\sl{H}\,||\,\sl{c}}$ at low temperatures, wheres the relation is reversed at high temperatures.
This is an important problem to be understood to further study the mechanism of the chiral soliton lattice state in Yb(Ni$_{1-x}$Cu$_x$)$_{3}$Al$_{9}$.\par
In order to clarify the mechanism of the change in easy axis of magnetization, we measured the magnetization processes of Yb(Ni$_{1-x}$Cu$_x$)$_{3}$Al$_{9}$ up to 14.5\,T while changing the field direction by 15 degrees. 
We also performed an analysis by introducing a crystalline electric field\,(CEF) level scheme.

\section{Experimental}
Single crystalline samples of Yb(Ni$_{1-x}$Cu$_x$)$_{3}$Al$_{9}$\,($x$$\leq$0.06) were prepared by Al self-flux method\cite{4}.
The starting materials of Yb, Ni, Cu, Al in molar ratio of 1\,: 3 (1--$x$)\,: 3$x$\,: 30\,($x$ = 0.0,\,0.1,\,0.2,\,0.3) were put in an alumina tube container and were sealed in a quartz tube.
The temperature was raised to 1000\,$^\circ$C, kept for 12 h, and was lowered to 700\,$^\circ$C at a rate of 2\,$^\circ$C per hour.
Excess Al-flux was removed with a centrifuge and the remaining Al was removed by etching with NaOH.
Magnetization was measured by an extraction method using a superconducting cryomagnet.

\section{Results and Discussion}
\begin{figure}[hbpt]
\includegraphics[width=15cm]{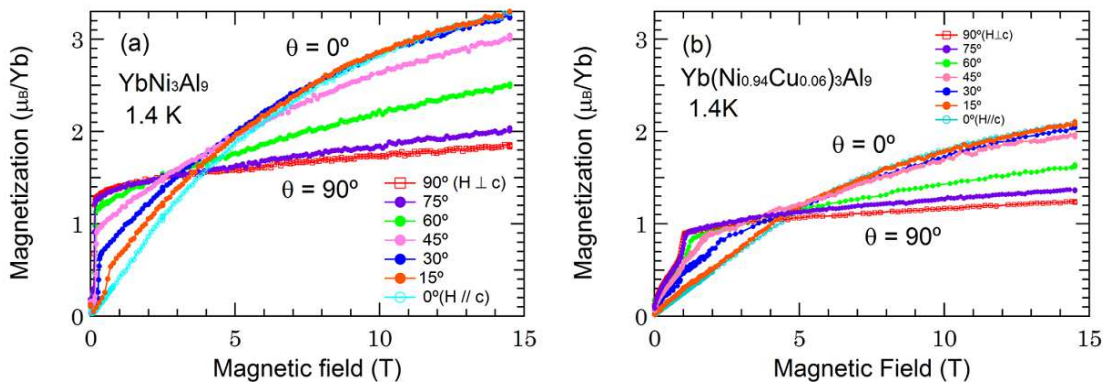}
\caption{Magnetization curves of Yb(Ni$_{1-x}$Cu$_x$)$_{3}$Al$_{9}$ for $x$\,=\,0\,(a) and $x$\,=\,0.06\,(b) at 1.4\,K. The sample was rotated by a step of 15$^\circ$ with respect to the field direction}
\label{MH1p4K}
\end{figure}

Figure\,\ref{MH1p4K} shows the $\sl{M}$($\sl{H}$) curves of Yb(Ni$_{1-x}$Cu$_x$)$_{3}$Al$_{9}$ for $x$\,=\,0 and 0.06 at the lowest temperature of 1.4\,K.
The measurement was performed by rotating the sample by a step of 15$^\circ$ with respect to the vertical field direction.
For $\sl{H}$\,$\perp$\,$\sl{c}$\,\,($\theta$\,=\,90$^\circ$), after the transition to the ferromagnetic state at 0.1\,T for $x$\,=\,0 and 1.0\,T for $x$\,=\,0.06, the magnetization hardly changes with increasing the field up to 14.5\,T.
However, when the field is applied along the $\sl{c}$-axis\,\,($\theta$\,=\,0$^\circ$), $\sl{M}$($\sl{H}$) gradually increases with increasing the field and overtakes the magnetization for $\sl{H}$\,$\perp$\,$\sl{c}$ at around 4\,T.
Above 4\,T, $\sl{M}$($\sl{H}$) for $\sl{H}$\,$||$\,$\sl{c}$ continues to increase, and finally at 14.5\,T, it becomes twice as large as that for $\sl{H}$\,$\perp$\,$\sl{c}$.
This overtaking behavior is commonly observed for other Cu concentrations as shown in Fig.\ref{MH1p4K}\,(b) for $x$\,=\,0.06 although the overall magnetization value is gradually reduced with increasing $x$.

\begin{figure}[hbpt]
\includegraphics[width=15cm]{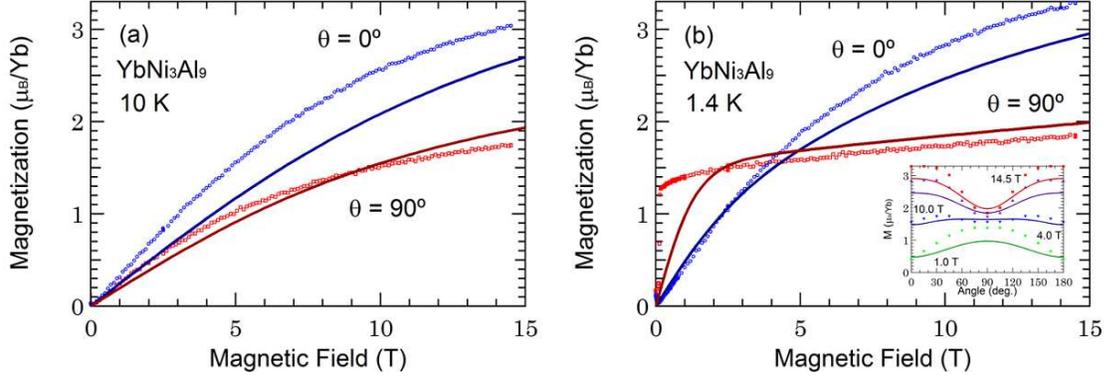}
\caption{Magnetization curves of YbNi$_3$Al$_9$ at 10\,K\,(a) and 1.4\,K\,(b). The solid lines are the calculated magnetization curves considering the single-site CEF effect. The inset shows the angular dependence of magnetization at 1.4\,K and the calculated magnetization curves.}
\label{MH_ExCalc}
\end{figure}

Figure\,\ref{MH_ExCalc}\,(a) shows the $\sl{M}$($\sl{H}$) curves of YbNi$_3$Al$_9$ at 10\,K in the paramagnetic phase.
We see that $\sl{M}$($\sl{H}$) for $\sl{H}$\,$||$\,$\sl{c}$ is about twice as large as that for $\sl{H}$\,$\perp$\,$\sl{c}$ in the whole field region.
This is consistent with the $\sl{M}$($\sl{H}$) curves at 1.4\,K in the high field region but is different from those in the low field region below the overtaking field of 4\,T.
Therefore, by combining the information on magnetic anisotropy observed in $\chi$($\sl{T}$), we may conclude that the $\sl{c}$-axis is the easy axis of magnetization at high temperatures or at high fields, whereas it becomes the hard axis at low fields at low temperatures.

\setlength\intextsep{10pt}

To analyze these results, let us introduce a CEF for the Yb$^{3+}$ ion. 
The $\sl{J}$\,=\,7/2 state splits into four Kramers' doublets.
It is estimated from specific heat measurement that the first excited state is located at $\Delta$$_1$\,=\,47\,K\cite{2}.
We searched for the CEF parameters so that the magnetization curves can be reproduced and the $\Delta$$_1$ is consistent with the reported value.
The CEF parameters satisfying these conditions are summarized in Table\,\ref{t1}. 
The energy-level scheme $\Delta$$_0$--$\Delta$$_1$--$\Delta$$_2$--$\Delta$$_3$ obtained from these parameters is 0--49.9--64.6--76.3\,K.
The calculated $\sl{M}$($\sl{H}$) curves at 1.4\,K and at 10\,K, taking into account the single site CEF effect only, are shown by the solid lines in Fig.\,\ref{MH_ExCalc}.
At 1.4\,K, the overtaking in the $\sl{M}$($\sl{H}$) curves is well reproduced as well as the magnetization values at high fields.
The discrepancy at low fields below 2\,T is due to the cooperative ordering phenomena which is not taken into account in the present calculation.\par
The CEF eigenfunctions in our model are written as

\setlength\intextsep{0pt}
\begin{table}
\begin{center}
\caption{CEF parameters used in the analysis to reproduce the magnetization curves.}
\label{t1}
\begin{tabular}{llllll}
\hline
$\sl{B}$$_{20}$ &$\sl{B}$$_{40}$ &$\sl{B}$$_{43}$ &$\sl{B}$$_{60}$ &$\sl{B}$$_{63}$ &$\sl{B}$$_{66}$ \\
-0.6 & -0.001 & 0 & 0.001 & 0 & 0.04 \\
\hline
\end{tabular}
\end{center}
\end{table}

\begin{eqnarray}
&\ket{0}&=\,-0.6862\ket{\pm5/2}+0.7274\ket{\mp7/2}\\
&\ket{1}&=\,\ket{\pm1/2}\\
&\ket{2}&=\,\ket{\pm3/2}\\
&\ket{3}&=\,+0.7274\ket{\pm5/2}+0.6862\ket{\mp7/2}
\end{eqnarray}

The overtaking in $\sl{M}$($\sl{H}$) is caused by the contribution from the $\sl{B}$$_{66}$ term, which connects the $\ket{\pm5/2}$ and the $\ket{\mp7/2}$ states.
Since the $\sl{B}$$_{20}$, $\sl{B}$$_{40}$ and $\sl{B}$$_{60}$ terms give rise to a definite anisotropy along or perpendicular to the $\sl{c}$-axis and cannot be associated with the overtaking behavior. 
The mixing between the $\ket{\pm5/2}$ and the $\ket{\mp7/2}$ states is important.
The $\sl{B}_{43}$ and $\sl{B}_{63}$ parameters, which generally arises in the site symmetry 3 of the rare-earth ion in the space group $\sl{R}32$, were found to be unimportant to explain the $\sl{M}$($\sl{H}$) curves.
They may be neglected.
This shows that the almost equilateral-triangular-prism shaped configuration of the surrounding atoms of Ni and Al around Yb is close to perfect.
In such a case the $\sl{B}_{43}$ and $\sl{B}_{63}$ terms vanish.\par
The mechanism of the magnetization process is understood by writing the matrix elements of $\sl{J}_x$ and $\sl{J}_z$ :

\[
 \sl{J}_x = \left(
    \begin{array}{cccc}
      \pm1.32 & 0 & 1.19 & \pm0.077 \\
      0 & \pm2.0 & \pm1.93 & 0 \\
      1.19 & \pm1.93 & 0 & 1.26 \\
      \pm0.077 & 0 & 1.26 & \pm 1.32
    \end{array}
  \right)
,\,\,
  \sl{J}_z = \left(
    \begin{array}{cccc}
      \pm0.67 & 0 & 0 & 2.99 \\
      0 & \pm0.5 & 0 & 0 \\
      0 & 0 & \pm1.5 & 0 \\
      2.99 & 0 & 0 & \pm 0.32
    \end{array}
  \right).
\]

\noindent
We see that the ground state has a larger moment along the $x$ direction ($\perp\sl{c}$) than the moment along the $z$ direction ($||\,\sl{c}$).
This leads the magnetic moments to be induced more in the $\sl{c}$-plane at low magnetic fields and at low temperatures and also to order in the $\sl{c}$-plane. 
However, at high fields or at high temperatures, the magnetic moments are more induced for $\sl{H}$\,$||$\,$\sl{c}$.
This is due to the Van-Vleck magnetization arising from the large off-diagonal matrix element $\bra{0}\sl{J}_z\ket{3}$.
Since $\ket{0}$ and $\ket{3}$ are composed of $\ket{\pm5/2}$ and $\ket{\pm7/2}$ states, these states can induce large moments along the $\sl{c}$-axis at high fields or at high temperatures.
Although the present analysis successfully explains the change in the easy axis at around 4\,T, the reduction in the magnetization and the magnetic anisotropy by the Cu substitution as shown in Fig.\,\ref{MH1p4K}\,(b) cannot be explained.
It might be associated with the change in carrier concentration, which may affect the CEF levels through a possible change in hybridization between 4$\sl{f}$ and conduction electrons.

\setlength\intextsep{10pt}
\begin{figure}[hbpt]
\includegraphics[width=15cm]{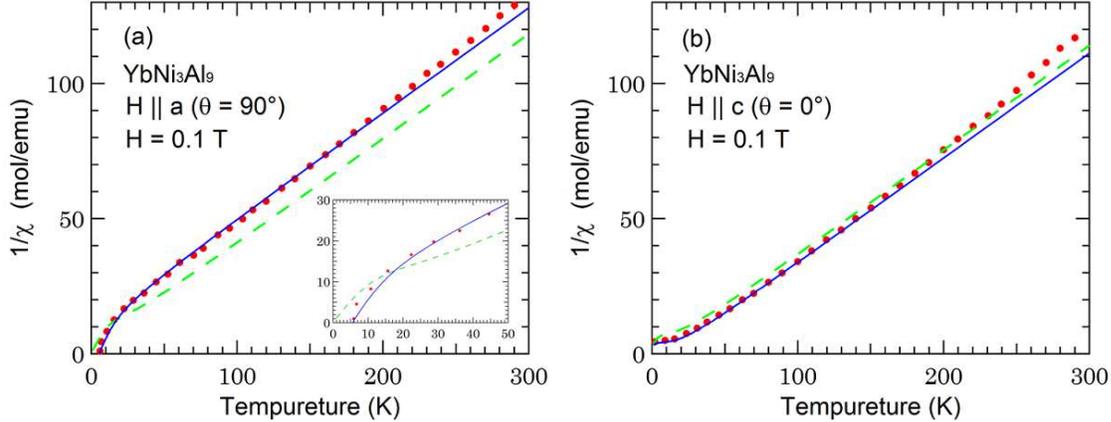}
\caption{Temperature dependence of inverse magnetic susceptibility of YbNi$_3$Al$_9$ for $\sl{H}$\,$||$\,$\sl{a}$\,(a) and $\sl{H}$\,$||$\,$\sl{c}$\,(b)\cite{2}. The solid lines are the calculated curves by considering the orbital dependent exchange in addition to the CEF effect. The dashed lines are the calculations by considering the CEF effect only. The inset shows the inverse magnetic susceptibility at low temperatures.}
\label{chiinv_exchange}
\end{figure}

Next, we discuss the temperature dependence of magnetic susceptibilities $\chi_a$($\sl{T}$) and $\chi_c$($\sl{T}$) reported in Ref.\,\citen{2} based on the present CEF model.
The experimental data and the calculated 1/$\chi$($\sl{T}$) from the present CEF model are compared in Fig.\,\ref{chiinv_exchange} as shown by the dashed lines.
For $\sl{H}$\,$||$\,$\sl{c}$ the data are well explained by the CEF effect only.
By shifting the dashed line slightly downwards, assuming a ferromagnetic exchange, the fit is expected to be improved.
However, for $\sl{H}$\,$||$\,$\sl{a}$, a uniform shift of the dashed curve upwards, assuming an antiferromagnetic exchange, cannot fit the data satisfactorily.
The downward curvature below 30\,K cannot be reproduced by simply shifting the dashed curve.
The slope of 1/$\chi$($\sl{T}$) at high temperatures are also different between the experiment and the calculation.
These results suggest that the exchange is not uniform and is dependent on the CEF states.\par
We introduce orbital dependent exchange interactions, where the molecular fields originating from different orbitals have different effects\cite{12}.

\begin{eqnarray}
\langle\mu_i\rangle&=&\,\chi_{i}^{(0)}\Bigl(\sl{H}\,+\,\sum_{j}\lambda_{ij}\langle\mu_j\rangle\,+\,\sum_{j,k}\lambda_{i[jk]}\langle\mu_{[jk]}\rangle\Bigr)\,,\\
\langle\mu_{[ij]}\rangle&=&\,\chi_{[ij]}^{(0)}\Bigl(\sl{H}\,+\,\sum_{k}\lambda_{k[ij]}\langle\mu_k\rangle\,+\,\sum_{k,l}\lambda_{[ij][kl]}\langle\mu_{[kl]}\rangle\Bigr)\,,\\
\langle\mu\rangle&=&\,\sum_{i}\langle\mu_i\rangle\,+\,\sum_{i,j}\langle\mu_{[ij]}\rangle\,,
\end{eqnarray}
\noindent
where $\langle\mu_i\rangle$ represents the magnetic moment induced in the CEF state $\ket{i}$ and $\langle\mu_{[ij]}\rangle$ the Van-Vleck moment induced between $\ket{i}$ and $\ket{j}$.
$\chi_i^{(0)}$ is the Curie susceptibility of the CEF state $\ket{i}$ and $\chi_{[ij]}^{(0)}$ the Van-Vleck susceptibility between the CEF states $\ket{i}$ and $\ket{j}$.
$\lambda_{ij}$, $\lambda_{i[jk]}$ and $\lambda_{[ij][kl]}$ are the mean-field exchange constants between $\langle\mu_i\rangle$ and $\langle\mu_j\rangle$,\, $\langle\mu_i\rangle$ and $\langle\mu_{[jk]}\rangle$,\, and between $\langle\mu_{[ij]}\rangle$ and $\langle\mu_{[kl]}\rangle$,\, respectively.
Since this original expression has too many parameters, we simplify the above formulas as the following:

\begin{eqnarray}
\langle\mu_{\rm{c}}\rangle&=&\,\chi_{\rm{c}}^{(0)}\Bigl(\sl{H}\,+\,\lambda_{\rm{cc}}\langle\mu_{\rm{c}}\rangle\,+\,\lambda_{\rm{cv}}\langle\mu_{\rm{v}}\rangle\Bigr)\,,\\
\langle\mu_{\rm{v}}\rangle&=&\,\chi_{\rm{v}}^{(0)}\Bigl(\sl{H}\,+\,\lambda_{\rm{cv}}\langle\mu_{\rm{c}}\rangle\,+\,\lambda_{\rm{vv}}\langle\mu_{\rm{v}}\rangle\Bigr)\,,\\
\langle\mu\rangle&=&\,\langle\mu_{\rm{c}}\rangle\,+\,\langle\mu_{\rm{v}}\rangle\,,
\end{eqnarray}

\noindent
where $\langle\mu_{\rm{c}}\rangle$ and $\langle\mu_{\rm{v}}\rangle$ represent the Curie and Van-Vleck moments, respectively. \par
The calculated 1/$\chi$($\sl{T}$) curves shown by the solid lines in Fig.\,\ref{chiinv_exchange} are obtained by assuming the following exchange parameters: $\lambda_{\rm{cc}(a)}$\,=\,6, $\lambda_{\rm{cv}(a)}$\,=\,$-$6, $\lambda_{\rm{vv}(a)}$\,=\,$-$20, $\lambda_{\rm{cc}(c)}$\,=\,$-$5, $\lambda_{\rm{cv}(c)}$\,=\,0, $\lambda_{\rm{vv}(c)}$\,=\,4\,\\(mol/emu). The experimental data, especially the $\sl{T}$-dependence of 1/$\chi$ for $\sl{H}$\,$||$\,a, are well explained by introducing the orbital dependent exchange parameters. The deviation of 1/$\chi$ from the linear Curie-Weiss law below 30\,K is caused not only by the CEF effect but also by the increasing contribution from the Curie-term susceptibility of  $\bra{0}\sl{J}_x\ket{0}$ with ferromagnetic interaction. The larger slope of 1/$\chi$ in the high temperature region than that expected from the CEF effect is due to the increasing contribution from the Van-Vleck susceptibility of $\bra{0}\sl{J}_x\ket{3}$ with antiferromagnetic interaction. The discrepancy between the signs of $\lambda_{(a)}$ and $\lambda_{(c)}$ needs to be improved. 

\section{Conclusion}
From the analyses of magnetization and magnetic susceptibility, we propose that the unusual magnetization process observed in rare-earth chiral magnet Yb(Ni$_{1-x}$Cu$_x$)$_{3}$Al$_{9}$, where the magnetization along the hard axis\,($||$\,$\sl{c}$) at low temperatures and at low fields overtakes the easy axis magnetization above $\sim$ 4\,T, is caused by the off-diagonal Van-Vleck magnetization between the ground state and the excited state of the CEF split levels. 
The $\sl{B}_{66}$-term of the CEF, which connects the $\ket{\pm5/2}$ and the $\ket{\mp7/2}$ states, is responsible for this behavior. The temperature dependence of magnetic susceptibility was also explained successfully by considering an orbital dependent exchange interaction in addition to the CEF effect.
It is expected in the future that the mechanism of CSL formation will be elucidated in more detail by taking into account the CEF wavefunctions obtained from the present analysis.

\section{Acknowledgement}
This work was supported by JSPS KAKENHI Grant number 18K187370A.
Magnetic susceptibility measurement using the MPMS was performed at N-BARD, Hiroshima University.

\end{document}